\documentclass[aps, prd, twocolumn, showpacs, superscriptaddress, groupedaddress]{revtex4} 
\usepackage{graphicx}    
\usepackage{amssymb}
\usepackage{subfigure}
\usepackage{float}
\usepackage[pagebackref=false, colorlinks=true]{hyperref}
\hypersetup{
linkcolor=blue,     % color of internal links
citecolor=blue,     % color of links to bibliography
urlcolor=blue}   
\usepackage{soul}
\usepackage{amsmath}
\begin{document}
\title{Causal Structure of Singularity in non-spherical Gravitational Collapse}
\author{Dipanjan Dey}
\email{dipanjandey.icc@charusat.ac.in}
\affiliation{International Center for Cosmology, Charusat University, Anand 388421, Gujarat, India}
\author{Pankaj S. Joshi}
\email{psjprovost@charusat.ac.in}
\affiliation{International Center for Cosmology, Charusat University, Anand 388421, Gujarat, India}
\author{Karim Mosani}
\email{kmosani2014@gmail.com}
\affiliation{BITS Pilani K.K. Birla Goa Campus, Sancoale, Goa 403726, India}
\author{Vitalii Vertogradov}
\email{vdvertogradov@gmail.com}
\affiliation{Department of Theoretical Physics and Astronomy, Herzen State Pedagogical University of Russia, Moika 48, Saint-Petersburg, 191186, Russia}
\affiliation{The SAO RAS, Pulkovskoe Shosse 65, 196140 St. Petersburg, Russia}

\date{\today}

\begin{abstract}
We investigate here the final state of gravitational collapse of a non-spherical and non-marginally bound dust cloud as modelled by the Szekeres spacetime. We show that a  directionally globally  naked singularity can be formed in this case near the collapsing cloud boundary, and not at its geometric center as is typically the case for a spherical gravitational collapse. This is a strong curvature naked singularity in the sense of Tipler criterion on gravitational strength. The null geodesics escaping from the singularity would be less scattered in this case in certain directions since the singularity is close to the boundary of the cloud as is the case in the current scenario. The physical implications are pointed out.
\bigskip
\end{abstract}
\maketitle

The visibility or otherwise of spacetime singularities forming due to gravitational collapse phenomenon in general relativity has been a subject of topical research today \cite{CRS1, CRS2, CRS3, CSR4, PSJ1, PSJ2, CSR5, MG1, MG2, TPS, Giambo, Giambo2, PSJ3}. The key reason for this interest is, if the singularities forming in collapse are visible to faraway observers in the universe as opposed to being hidden within an event horizon of a black hole,
we will be able to see then the physical signatures of the possible quantum gravity effects that may occur near such singularities due to the ultra-strong gravity fields present there.

It was shown earlier that singularities formed due to gravitational collapse of a dust cloud can be visible to an asymptotic observer in principle \cite{Deshingkar, Giambo2, Mosani}. These are called globally naked singularities as opposed to locally naked ones where the outgoing light and particle trajectories from the singularity again fall back within the event horizon later. 
Certain observational signatures that could distinguish between hidden and naked singularities, like the shadows it forms \cite{Shaikh}, the timelike and null trajectories in its vicinity \cite{Bambhaniya, Dey}, the accretion disc properties \cite{Shaikh2}, and the gravitational lensing \cite{Bambi, Dey2} have been studied in detail. The singularity theorems prove that under generic conditions, singularities would form in gravitational collapse as well as in cosmology \cite{Penrose, Hawking}. One of the theorem's interpretations is that general relativity predicts its own breakdown since proper physical variables must not be allowed to blow up indefinitely. Hence, existence of singularities puts a question on the validity of general relativity in the strong gravity regime. As a result, an improved theory of gravity closer to the approach of singularities is presumed \cite{Wheeler, Hawking2, Bergmann}. As an implication a viewpoint is suggested, namely that the physics near the singularities should be investigated further, since it may guide us into what one could expect from improved theories of gravity such as quantum gravity, applicable for strong gravity regimes \cite{Misner, Burko}. 

A Spherically symmetric collapsing cloud forms a singularity at the center which may be hidden in a black hole or it is a visible naked singularity.
A relevant question here is, even if the singularity is naked, whether it would be visible to faraway observers in the universe in physical reality. For example, in spherical collapse models,
even when the singularity is globally naked allowing the timelike and null geodesics to escape away from the collapsing cloud, the singularity occurs only at the center of the collapsing matter cloud. Such a singularity, even when it is visible faraway in principle, may not be able perhaps to radiate away energy, being embedded within very high density regions of the collapsing star. Thus the physical implications of such a naked singularity would need a more detailed investigation.

Even if the singularity of collapse is not hidden to faraway observers, one could argue that the information about the extremely strong gravity region carried by the outgoing null geodesics can be opaque or get distorted  because of the scattering of the null geodesic due to the collapsing matter surrounding the singularity. The spherical symmetry is, however, a strict presupossition and analysing more general solutions of the Einstiens field equation would be of considerable interest.
In other words, it is important to ask whether the same scenario persits in collapse models that are not exactly spherical, or which represent small perturbations from spherical symmetry.

Existence of strong (in the sense of Tipler \cite{Tipler, Clarke}), naked singularities formed due to quasi-spherical collapsing clouds governed by Szekeres spacetimes was studied previously by  Joshi and Krolak \cite{Joshi}. It was shown that the criteria for such singularity to be  strong naked is the same as in the case of spherical symmetry. Globally visibile singularity was shown to exist in collapse of such non-spherical marginally bound cloud by Deshingkar \textit{et.al.} \cite{Deshingkar}.

What we need to inquire and examine is whether the causal structure of the naked singularity remains the same, i.e. embedded in the interior of the matter cloud, or whether it exibits other novel causal features. In particular, 
the distortion of the information can be reduced or avoided if the singularity is formed near the boundary of the collapsing cloud. Closer the singularity is to the boundary, lesser will be the scattering of the outgoing singular null geodesics in certain directions. We show here that such collapsing spacetime solutions, which can end up in a singularity which is not at the geometric center, can be obtained from the Szekeres solution \cite{Szekeres}.

Marginally bound collapse is again a special case, so to offer generality we investigate here a non-marginally bound non-spherical gravitational collapse.
We show that for a suitable choice of free functons, arising due to available degrees of freedom in the Einstein's field equations, one could achieve a scenario wherein the globally visible singularity is formed not at the geometric center but closer to the boundary of the collapsing cloud. 

It thus turns out that introducing asphericity in collapse can radically alter the causal structure of the naked singularity, in that its visibility can be greatly enhanced. This indicates that small perturbations from spherical symmetry are important to consider in order to examine the physical implications of naked singularities.

Using the units in which $c=8\pi G=1$, the general Szekeres metric in the comoving coordinates is given by,
%\begin{equation}
 %   ds^2=-dt^2+M^{2}dr^2+N^{2}(dx^2+dy^2).
%\end{equation}
%The energy-momentum tensor of \textit{type I} matter field with vanishing pressure is written as $T^{\mu \nu}=\rho U^{\mu} U^{\nu}$, where $U^{\mu}$ are the components of the four velocity. We now consider a pair of conjugate coordinates defined as $\zeta=x+i y$ and $\Bar{\zeta}=x-i y$. The metric can then be rewritten in the $(t,r,\zeta,\Bar{\zeta})$ coordinates as
\begin{equation*}
    ds^2=-dt^2+M^{2}(t,r,\zeta,\Bar{\zeta})dr^2+N^{2}(t,r,\zeta,\Bar{\zeta})d\zeta d\Bar{\zeta},
\end{equation*}
where  $\zeta=x+i y$ and $\Bar{\zeta}=x-i y$ is a pair of conjugate coordinates, $N=S(t,r)/Q(r,\zeta,\Bar{\zeta})$ and $M=Q N'/\sqrt{1+f(r)}$, and $f>-1$ is the velocity function. $f$ greater than, equal to, and less than zero corresponds to bound, marginally bound, and unbound collapse respectively. Additionally, $N'\neq 0$. Here the subscripts prime and dot denotes the partial derivative with respect to $r$ and $t$ respectively. Also,
\begin{equation*}
    Q=a(r)\zeta \Bar{\zeta}+B(r)\zeta+\Bar{B}(r)\Bar{\zeta}+c(r),
\end{equation*}
where $a$ and $c$ are real, and $B$ is a complex function having the relation $ac-B\Bar{B}=\delta/4$, where $\delta=0,\pm 1$. The energy-momentum tensor of \textit{type I} matter field with vanishing pressure is written as $T^{\mu \nu}=\rho U^{\mu} U^{\nu}$, where $U^{\mu}$ are the components of the four velocity.
%%%%%%%%%%%%%%%%%%%%%%%%%%%%%%%%%%%%%%%%%%%%%%%%%%%%%%%%%%%%%%%%%%%%%%%%%%%%%%%%%%%%%%%%%%%%%%%%%%%%
\begin{figure*}\label{fig1}
{\includegraphics[scale=0.5]{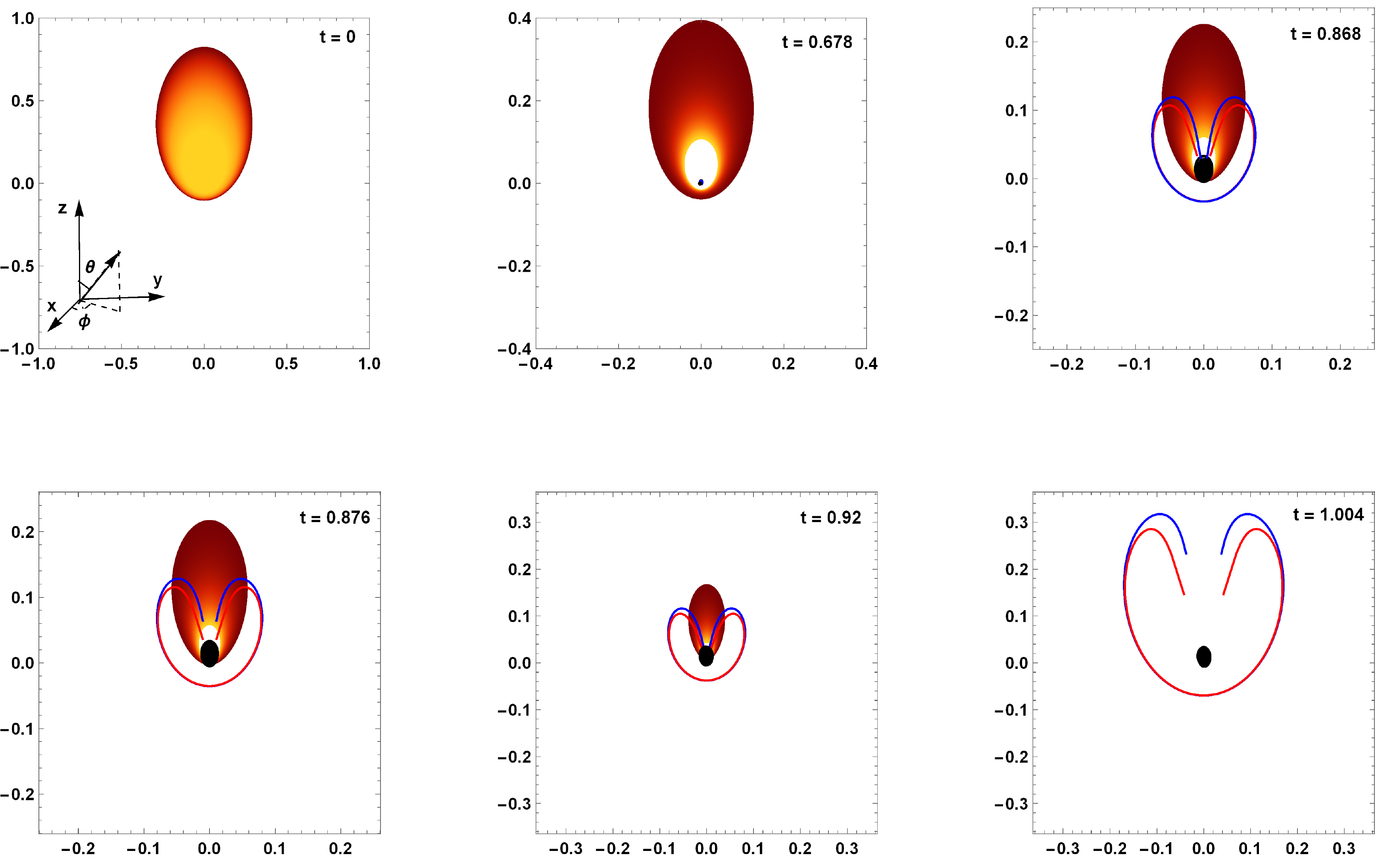}}
%\hspace{0.2cm}
%\subfigure[]
\caption{Cross-section along fixed $\phi=\pi/2$ of the evolution of the collapsing aspherical dust cloud, and the global causal structure of the singularity thus formed, is depicted. The figures are symmetric with change in $\phi$. Here, $A=r^3-20r^6$, $f=-10^{-3}r^2$ and $Q=\text{exp}(r) \tan^2{\left (\frac{\theta}{2}\right)}+\frac{\text{exp}(-r)}{4}$. The singularity forms at $t=2/3$. Evolution of event horizon depends on $\theta$. The event horizon touches the boundary of the collapsing cloud first at $\theta=\pi$ at $t=0.868$. The aspherical wavefronts of singular null geodesics (red and blue colored) escape the singularity, thereby making it globally visible. However, part of these outgoing wavefronts which lies in the neighbourhood of $\theta=0$ gets trapped by the trapped surfaces and falls back to the singularity, hence making it only directionally globally naked in the neighbourhood around $\theta=\pi$.   }
\end{figure*}
%%%%%%%%%%%%%%%%%%%%%%%%%%%%%%%%%%%%%%%%%%%%%%%%%%%%%%%%%%%%%%%%%%%%%%%%%%%%%%%%%%%%%%%%%%%%%%%%%%%
%%%%%%%%%%%%%%%%%%%%%%%%%%%%%%%%%%%%%%%%%%%%%%%%%%%%%%%%%%%

%Choosing $\zeta=f_1(\theta)e^{i \phi}$ gives us
%\begin{equation}
%   d\zeta d\Bar{\zeta}=\left \vert \frac{\partial f_1}{\partial \theta}\right \vert^2d\theta^2+\left \vert \frac{\partial f_1}{\partial \phi} \right \vert^2 d\phi^2.
%\end{equation}
Choosing  $\zeta=\tan(\theta/2)e^{i \phi}$, and $B=0$, one can express $N^2d\zeta d\Bar{\zeta}$ as 
\begin{equation*}
    N^2d\zeta d\Bar{\zeta}=
    R^2(t,r,\theta) \left(d\theta^2 +\sin^2{\theta}d\phi^2\right),
\end{equation*}
where
\begin{equation*}
    R(t,r,\theta)=\frac{S(t,r)\sec^2(\frac{\theta}{2})}{2(a(r)\tan^2(\frac{\theta}{2})+c(r))}
\end{equation*}
is the physical radius of the collapsing cloud, which tells us the distance of a point $(t,r,\theta, \phi)$ from its center of mass. Note that $R$ is symmetric with respect to change in $\phi$. The density and pressure of the collapsing cloud, obtained from the Einstein's field equation, are expressed respectively as
\begin{equation}\label{efe}
    \rho(t,r,\theta)=\frac{QA'-3AQ'}{S^2(Q S'-S Q')},
\hspace{1cm}    p(t,r)=-\frac{\dot A}{S^2 \dot S},
\end{equation}
where
\begin{equation} \label{A}
    A=S(\dot S ^2-f).
\end{equation}
Assumption of zero pressure corresponds to $A$ being a function of only $r$, as is apparent from Eq.(\ref{efe}). 
%It should be noted that the density can also be written as $\rho(t,r,\theta)=F'/(R^2R')$, where $F$ is twice the mass inside a collapsing shell of constant radial coordinate $r$ at time $t$. 
Eq.(\ref{A}) can be integrated to obtain
\begin{equation}\label{timecurve}
    t-t_s(r)=-\frac{S^{\frac{3}{2}}\mathcal{G}(-fS/A)}{\sqrt{A}},
\end{equation}
 %%%%%%%%%%%%%%%%%%%%%%%%%%%%%%%%%%%%%%%%%%
%%%%%%%%%%%%%%%%%%%%%%%%%%%%%%%%%%%%%%%%%%%%%%%%%%%%%%%%%%%%%%%%%%%%%%%%%%%%%%%%%%%%%%%%%%%%%%%%%%%%
\begin{figure*}\label{fig1}
\begin{subfigure}
{\includegraphics[scale=0.5]{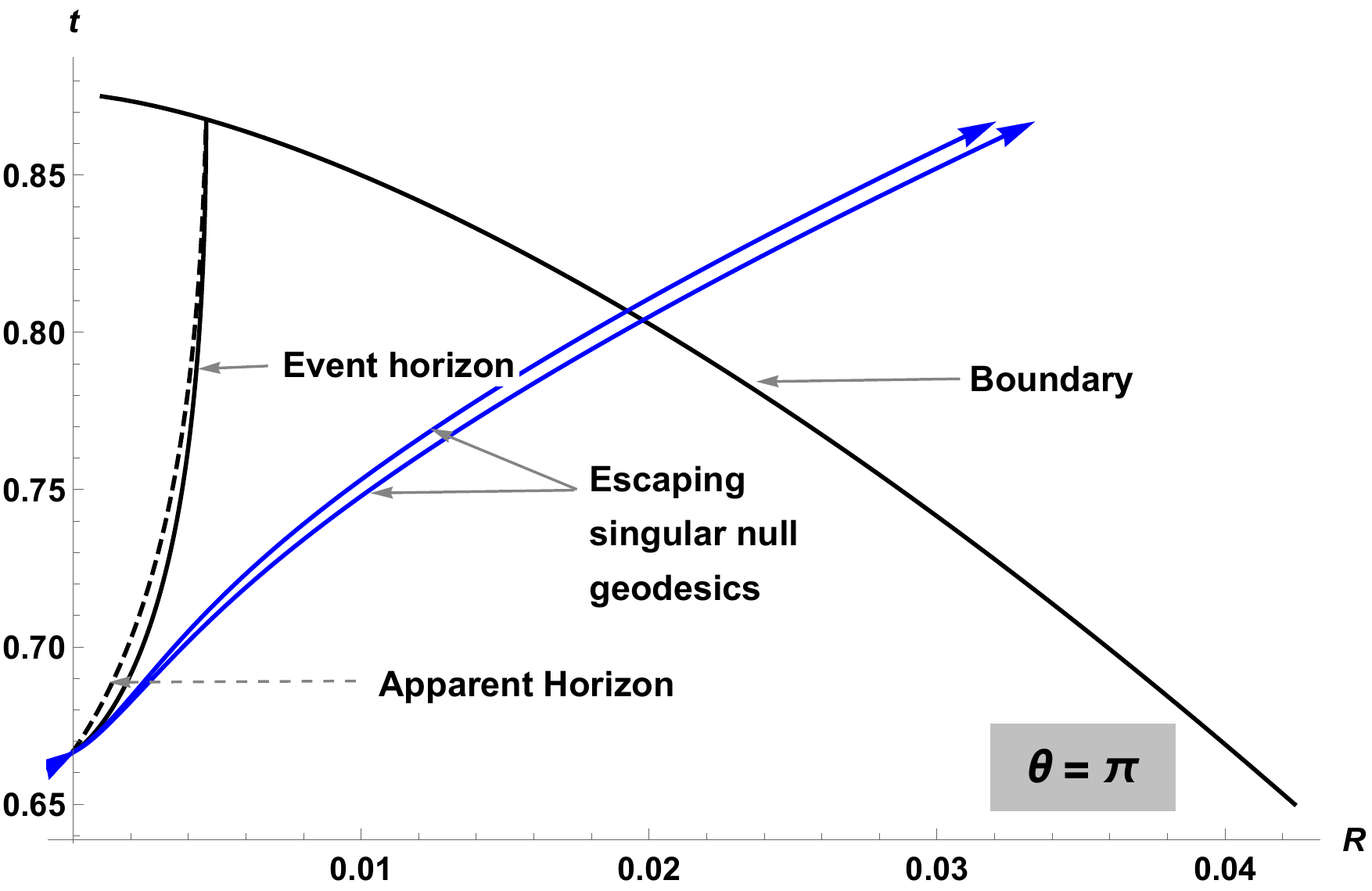}}{\includegraphics[scale=0.5]{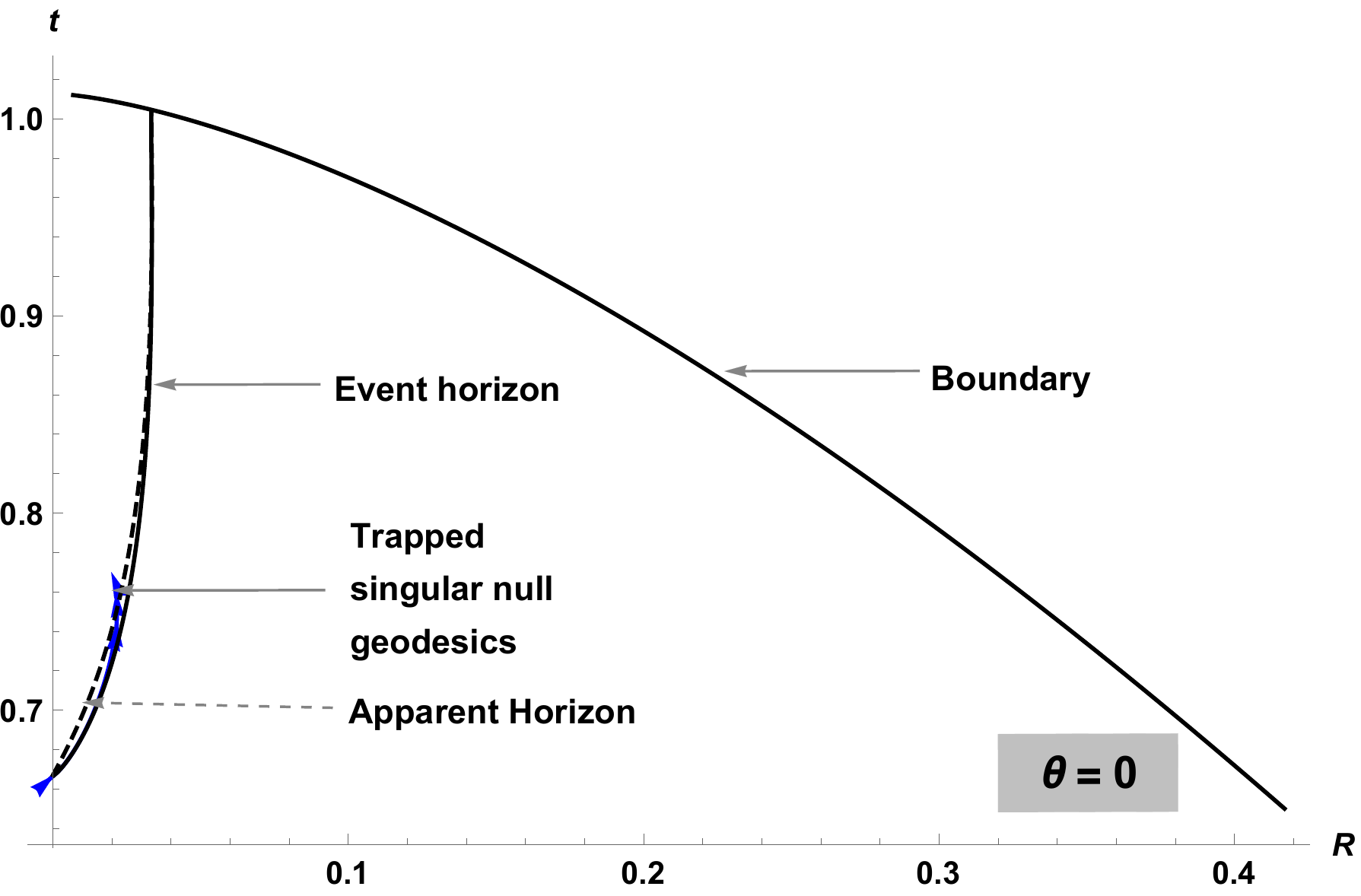}}
\end{subfigure}
%\hspace{0.2cm}
%\subfigure[]
\caption{Causal structure of the singularity formed as the end state of a bound (elliptic) collapsing aspherical dust cloud along different inclination angles $\theta=\pi$ and $\theta=0$.  The apparent horizon, event
horizon, and singular null geodesics are represented by dashed black curves, solid black curves, and solid blue curves, respectively. Here, $A=r^3-20r^6$, $f=-10^{-3}r^2$ and $Q=\text{exp}(r) \tan^2{\left (\frac{\theta}{2}\right)}+\frac{\text{exp}(-r)}{4}$. Escaping singular null geodesics can reach the boundary of the collapsing cloud at $\theta=\pi$, and these get trapped and fall back to the singularity at $\theta=0$. The singularity is thus directionally globally visible.}
\end{figure*}
%%%%%%%%%%%%%%%%%%%%%%%%%%%%%%%%%%%%%%%%%%%%%%%%%%%%%%%%%%%%%%%%%%%%%%%%%%%%%%%%%%%%%%%%%%%%%%%%%%%
where $\mathcal{G}(y)$ is defined as follows:
\begin{equation*}\label{G}
    \begin{split}
        & \mathcal{G}(y)= \left(\frac{\text{arcsin}\sqrt{y}}{y^{\frac{3}{2}}}-\frac{\sqrt{1-y}}{y} \right) \hspace{0.5cm} \text{for} \hspace{0.5cm} 0<y<1, \\
        & \mathcal{G}(y)=\frac{2}{3} \hspace{0.5cm} \text{for} \hspace{0.5cm} y=0, \\
        & \mathcal{G}(y)= \left(\frac{-\text{arcsinh}\sqrt{-y}}{(-y)^{\frac{3}{2}}}-\frac{\sqrt{1-y}}{y} \right) \hspace{0.5cm} \text{for}  -\infty<y<0. 
    \end{split}
\end{equation*}
Here, $t_s(r)$, is the time at which a collapsing shell of fixed radial coordinate $r$ becomes singular, and is given by
\begin{equation*}\label{singularitycurve}
    t_s(r)=\frac{S_0(r)^{\frac{3}{2}}}{\sqrt{A}}\mathcal{G}\left(-\frac{f S_0(r)}{A}\right),
\end{equation*}
where $S_0(r)=S(0,r)$. Consider the comoving radius corresponding to the boundary of the collapsing cloud $r_c$. In our aspherical collapsing model, the density vanishes at the boundary, hence $r_c=r_c(\theta)$. Therefore, from the above equation, $t_s(r_c)=t_s(r_c(\theta))$. Hence, the boundary of the cloud does not collapse to the singularity simultaneousy, but rather falls in the singularity at different times along different directions.  The apparent horizon, which is the boundary of trapped surfaces forming inside the collapsing cloud, is represented by vanishing $\Theta_l$, where
\begin{equation*}
    \Theta_l=h^{\mu \nu}\nabla_{\mu}l_{\nu}=\left (g^{\mu \nu}+\frac{l^{\mu}n^{\nu}+l^{\nu}n^{\mu}}{-l^{\alpha}n_{\alpha}}\right)\nabla_{\mu}l_{\nu}
\end{equation*}
is the expansion scalar of outgoing null geodesic congruence. Here, $h^{\mu \nu}$ is the transverse metric, and $l^{\alpha}$ and $n^{\alpha}$ are the tangents of the outgoing and incoming null geodesics respectively. We  obtain from here that along the apparent horizon curve, $A=S$. Using this along with Eq.(\ref{timecurve}), we obtain the apparent horizon curve as
\begin{equation}
    t_{AH}(r)= t_s(r)-A\mathcal{G}(-f).
\end{equation}
From this equation, it can be seen that $t_s(0)=t_{AH}(0)$, since $A(0)=0$, as demanded by the regularity conditions \cite{Joshi2}. 
%[HOW SPECIAL OR GENERIC ARE THESE VALUES CHOSEN? IN OTHER WORDS, IS THERE A PARAMETER RANGE FOR WHICH SAME OR SIMILAR RESULTS ARE OBTAINED, OR THIS IS JUST A SINGLE EXAMPLE?] 

The event horizon is a null surface and its evolution is described by the solution of the nul geodesic equation
\begin{equation}
\frac{dt}{dr}=M(t,r,\theta),    
\end{equation}
 satisfying the condition, $A=S$ at $r=r_c$. For a singularity to be visible globally in a certain constant $\theta$ direction, the event horizon at the center of mass should form not before, but at the time of formation of the singularity obtained due to the collapsing shell corresponding to $r=0$. If the event horizon at the center of mass forms before the singularity, in a given direction, even if the singular null geodesic escapes, it later gets trapped and falls back in, before reaching the collapsing boundary.
 %Additionally, the singularity should be a nodal point. This ensures that entire family of null geodesics escape from the singularity $(t_s(0),0)$ in the $(t,R)$ plane with fixed $\theta$ coordinate, thereby exposing the singularity globally, for infinite time.

As seen in Fig.(1), for suitable functions $A$, $f$ and $Q$ (freedom of choice possible due to the availability of three degrees of freedom in the given set of Einstein's equations), satisfying regularity conditions \cite{Joshi2}, a  directional globally visible singularity forms at $(0,0)$, which is away from the geometric center of the collapsing cloud, and is located near its boundary. Directional visibility is attributed to the fact that the event horizon starts evolving from the center at different times along different $\theta$ (also see Fig(2)). The null geodesics should escape from the region sufficiently close to the singularity to contain signatures of the strong gravity regions (mimicing possibly the quantum gravity effects); i.e. the difference in time of escape of the null singularity at $r=0$ and the time of formation of the singularity at $r=0$ should be of the order of the Planck time. The outermost aspherical wavefront of the null geodesic satisfying this criterion lies partly inside and partly outside the aspherical event horizon surface, intersecting it such that the locus of the points of intersection is a closed curve. This closed curve subtends a solid angle at the center $(0,0)$ such that the part of escaping singular null geodesic wavefront lying inside this solid angle gets trapped and the part outside escapes, thereby making the singular region visible, only if the asymptotic observer lies outside the region subtended by this solid angle.
It is worth mentioning that the aspherical collapse example which we have shown is not a small perturbation on the LTB metric. We know that in the case of spherical symmetry, there exists a non-zero measured set of initial parameters giving rise to such globally visible singularity  \cite{Joshi2}.
Similarly, one can show that there exists a non-zero measured set of initial parameters $A$ and $f$ giving rise to a directional globally visible singularity which is offcentric and close to the boundary of the collapsing cloud.

%[A RELEVANT QUESTION HERE IS, CAN THESE BE REGARDED AS "SMALL PERTURBATIONS" FROM SPHERICAL SYMMETRY? IS IT A SMALL CHANGE IN VALUES FROM THE LTB METRIC?] 
This directonally visible singularity is physically strong in the sense that at least along one null geodesic with affine parameter $\lambda$, with $\lambda=0$ at the singularity, the inequality $\lambda^2R_{ij}K^iK^j>0$ holds as $\lambda \to 0 $. Here $R_{ij}$ is the Ricci tensor, and $K^{\alpha}$ is the tangent of the outgoing null geodesics from the singularity. This condition ensures that the volume element formed by independent Jacobi fields along the geodesic vanishes as the geodesic terminates at the singularity. 

Once the matter cloud falls inside the event horizon, the exterior spacetime is aspherical, static, vacuum, and asymptotically flat. It was shown by \cite{Cruz, Price} using perturbation theory, that for a small deformation from spherical symmetry, a non-rotating, collapsing body radiates away the deformatons in the form of gravitational waves and assumes the shape of minimum curvature (as visualized by Misner \cite{Thorne}). Hence, the end state is possibly a spherical event horizon. For highly deformed spacetime, this may or may not be true \cite{Thorne}. 

In conclusion, we note that the singular null geodesics escaping radially in the neighbourhood around $\theta=\pi$ (Fig(1)) are less scattered due to interaction with the collapsing matter, than those in other directions. These less scattered geodesics  in a less distorted form may contain traces of gravity theory that governs the strong gravity regime. One might wonder if there is a scenario, a solution of the Einstein's field equation, wherein the singularity forms `at' the boundary (and not close to it) so that the original form of the escaping singular null geodesic is maintained with no distortion at all.

\textit{Acknowledgement:}
K.M. would like to acknowledge the Council of Scientific and Industrial Research (CSIR, India, Ref: 09/919(0031)/2017-EMR-1) for funding the work. V.V: The work was performed within the SAO RAS state assignment in the part ``Conducting Fundamental Science Research".

\end{document}